\documentclass[prd,aps,twocolumn,showpacs,preprintnumbers,amsmath,amssymb,nofootinbib]{revtex4}
\usepackage[dvips]{graphicx}
\usepackage{epsf}
\usepackage{amsmath}
\usepackage{amssymb}

\usepackage{graphicx}
\usepackage{dcolumn}
\usepackage{bm}
\def\be{\begin{equation}}
\def\ee{\end{equation}}
\def\bea{\begin{eqnarray}}
\def\eea{\end{eqnarray}}

\begin{document}

\title{Testing the Lorentz and CPT Symmetry with CMB polarizations and a non-relativistic Maxwell Theory}

\author{Yi-Fu Cai$^{1}$, Mingzhe Li$^{2,3}$ and Xinmin Zhang$^{1,4}$}

\affiliation{1) Institute of High Energy Physics, Chinese Academy
of Sciences, P.O. Box 918-4, Beijing 100049, P.R. China}

\affiliation{2) Department of Physics, Nanjing University, Nanjing
210093, P.R. China}

\affiliation{3) Joint Center for Particle, Nuclear Physics and
Cosmology, Nanjing University - Purple Mountain Observatory,
Nanjing 210093, P.R. China}

\affiliation{4) Theoretical Physics Center for Science Facilities
(TPCSF), Chinese Academy of Sciences, P.R. China}


\begin{abstract}
We present a model for a system involving a photon gauge field and
a scalar field at quantum criticality in the frame of a
Lifthitz-type non-relativistic Maxwell theory. We will show this
model gives rise to Lorentz and CPT violation which leads to a
frequency-dependent rotation of polarization plane of radiations,
and so leaves potential signals on the cosmic microwave background
temperature and polarization anisotropies.
\end{abstract}

\maketitle

\section{Introduction}

The Lorentz and the CPT (Charge conjugation-Parity-Time reversal)
symmetries are commonly assumed to be the most fundamental
elements in constructing the models of elementary particle
physics. Various attempts to break these symmetries have been
intensively discussed in the literature. Recently motivated by the
pioneering work in condensed matter physics\cite{Lifshitz},
Ho\v{r}ava has proposed a class of models of non-relativistic
quantum field theories \cite{Horava:2008jf, Horava:2008ih}, in
which the spatially higher derivative terms are introduced to
improve their ultraviolet behaviors. This theory, when applied to
the gravity sector, is power-counting renormalizable and hence
potentially ultraviolet (UV) complete\cite{Horava:2009uw},
although a strong coupling problem needs to be aware
of\cite{Charmousis:2009tc}. Moreover, a class of non-relativistic
Yang-Mills gauge theories in D+1 dimensions whose free-field limit
exhibits quantum critical behavior with gapless excitations and
dynamical critical exponent $z=2$ have been studied in Ref.
\cite{Horava:2008jf}. The class of these theories are expected to
be engineered from the D-brane configurations of string
theory\cite{Horava:2008ih}, and in the large-N limit, they have
weakly curved gravity duals. A general feature of this theory is
that, the action does not need the usual diffeomorphism invariance
of General Relativity, but has a fixed point in the infrared (IR)
limit which can recover the Lorentz invariance as an accident
symmetry. As a specific low energy scale presentation of these
theories, a non-relativistic Maxwell theory appears to be the most
potential one which might be detectable in experiments.

Probing for the violation of the Lorentz and the CPT symmetries is
an important approach to searching for new physics beyond the
standard model\cite{Carroll:1989vb, Harari:1992ea,
Colladay:1998fq, Jackiw:1999yp, Kostelecky:2003fs}. Up to now,
however the Lorentz and the $CPT$ symmetries have passed a number
of high precision experimental tests in the ground-based
laboratory and no definite signal of their violation has been
observed \cite{Lehnert:2006mn}. So, the present Lorentz and the
$CPT$ violating effects, if they exist, should be very small to be
amenable to the experimental limits. In the past several years,
there has been a lot of studies on the lorentz and CPT test with
the polarization of the cosmic microwave background radiation
(CMB). One of the examples is to consider a scalar field which
couples to the photon Chern-Simon term $\mathcal{L}_{\rm
int}\sim\partial_{\mu}{\phi}A_{\nu}\widetilde{F}^{\mu\nu}$\cite{Carroll:1989vb}.
This interaction is Lorentz invariant, however breaks dynamically
the CPT and lorentz symmetries during the evolution of the
universe when $\phi$ rolls down along a potential and $\dot{\phi}$
gets a non-vanishing value. The parity violation, resulted by the
rolling cosmological scalar\cite{Carroll:1989vb} as well as the
gravitational chirality\cite{Contaldi:2008yz}, leads to the
rotation of the CMB photon polarization when propagating in the
universe\cite{Harari:1992ea, Lue:1998mq, Feng:2004mq}, and the
current CMB data mildly indicates a non-zero value of this
rotation angle\cite{Feng:2006dp, Komatsu:2008hk, Xia:2008si,
Kostelecky:2008be}. Importantly in comparison to the laboratory
experiments, using the CMB polarization has been shown to be the
most powerful approach to testing the Lorentz and CPT symmetries
in the photon sector since the CMB photon has travelled a long
time (almost the age of the universe).

In this letter we study the model for a system consisting of a
scalar $\phi$ coupled to the photon field in a Lifshitz-type
non-relativistic Maxwell theory and discuss its implications for
CMB polarizations. Since the theory has different fixed points at
UV and IR limits respectively, our results show that the rotation
angle obtained in this model is quite different from what obtained
in the usual case. For high energy photons, the rotation angle is
frequency dependent and this dependence differs from that of
Faraday rotation, which is a new feature of this model. For low
energy photons, the rotation angle obtained in this model
coincides with the one for a dynamical scalar field coupled to the
Chern-Simons current in the $3+1$ dimensional spacetime.

This letter is organized as follows. In Section II we review very
briefly on the CMB polarization and its connection to the CPT
test. In Section III we review also very briefly on the
construction of a Lifthitz-type theory. In Section IV we present
in detail the model with $z=2$. In Section V we study the
implications of our model in CMB and calculate the rotation angle
of the CMB photon. 
Section VI is our conclusion and discussions.

\section{CMB polarization constraints on CPT violation}

In this section we review very briefly how to use the CMB
information to constrain the CPT violation. In general it is
convenient to use the Stokes parameters to study the CMB
polarization. Considering a monochromatic electromagnetic wave of
frequency $\omega$ propagating in the positive direction of $z$
axis, which is described by an electric field $\vec{E}$, the
Stokes parameters are defined as
\begin{eqnarray}
I &\equiv& \langle E_1E_1^*\rangle+\langle E_2E_2^*\rangle~, \nonumber\\
Q &\equiv& \langle E_1E_1^*\rangle-\langle E_2E_2^*\rangle~, \nonumber\\
U &\equiv& \langle E_1E_2^*\rangle+\langle E_1^*E_2\rangle~, \nonumber\\
V &\equiv& i[\langle E_1E_2^*\rangle-\langle E_1^*E_2\rangle]~,
\end{eqnarray}
where the notation $\langle\rangle$ denotes the time average.

Among these parameters, $Q$ and $U$ reflect the CMB polarization
which can be decomposed into a gradient-like ($E$) and a curl-like
($B$) component. For the standard theory of CMB, the $TB$ and $EB$
cross correlations vanish. However, with the presence of CPT
violation by the interaction terms mentioned above, we could
observe a rotation angle $\Delta\chi$ on the polarization vector
of a photon\cite{Lue:1998mq, Li:2008tma}.

In the following we study how the rotation angle affects the CMB
power spectra. In a flat FRW universe, we can expand the
temperature and polarization anisotropies in terms of
spin-weighted harmonic functions,
\begin{eqnarray}
T(\hat{\mathbf{n}}) &=&
\sum_{lm}c_{T,lm}Y_{lm}(\hat{\mathbf{n}})~,\nonumber\\
(Q \pm iU)(\hat{\mathbf{n}}) &=&
\sum_{lm}c_{\pm2,lm\pm2}Y_{lm}(\hat{\mathbf{n}})~,
\end{eqnarray}
where $\hat{\mathbf{n}}$ is the unit of the propagation direction,
and $Y_{lm}$ is the spherical harmonic function. One can study the
gradient-like component and curl-like component by introducing a
group of linear combinations of the coefficients $c_{-2,lm}$ and
$c_{+2,lm}$ as follows,
\begin{eqnarray}
c_{E,lm}=-\frac{1}{2}(c_{-2,lm}+c_{+2,lm})~,\nonumber\\
c_{B,lm}=-\frac{i}{2}(c_{-2,lm}-c_{+2,lm})~.
\end{eqnarray}

\begin{figure}[htbp]
\includegraphics[scale=0.4]{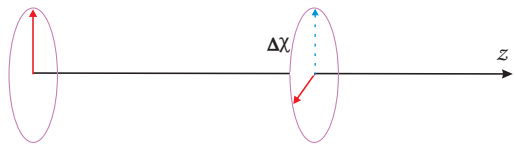}
\caption{A sketch plot of the generation of the rotation angle
$\Delta\chi$. An electromagnetic wave is propagating along the $z$
axis which is the horizontal one. The red arrow denotes the
polarization direction. After a period of the propagation, its
direction deviates from the original one.} \label{fig:sketch}
\end{figure}

With the assumption of statistical isotropy, all the CMB power
spectra can be defined as the following expressions,
\begin{eqnarray}
\langle c^*_{J',l'm'}c_{J,lm}
\rangle=C_l^{J'J}\delta_{l'l}\delta_{m'm}~,
\end{eqnarray}
where $J$ denotes the temperature $T$ and the $E$ and $B$ modes of
the polarization field respectively. To neglect the local
fluctuations, the CMB photons travel through the universe by
rotating their polarizations synchronously. Then we obtain the
rotation relations of the power spectra as follows,
\begin{eqnarray}
C_l^{TT,obs} &=& C_l^{TT}~,\\
C_l^{TE,obs} &=& C_l^{TE}\cos(2\Delta\chi)~,\\
C_l^{EE,obs} &=&
C_l^{EE}\cos^2(2\Delta\chi)+C_l^{BB}\sin^2(2\Delta\chi)~,\\
C_l^{BB,obs} &=&
C_l^{EE}\sin^2(2\Delta\chi)+C_l^{BB}\cos^2(2\Delta\chi)~.
\end{eqnarray}
Moreover, the $TB$ and $EB$ correlations do not vanish,
\begin{eqnarray}
C_l^{TB,obs} &=& C_l^{TE}\sin(2\Delta\chi)~,\\
C_l^{EB,obs} &=& \frac{1}{2}(C_l^{EE}-C_l^{BB})\sin(4\Delta\chi)~,
\end{eqnarray}
and thus the polarization surface can be rotated as sketch in
Figure \ref{fig:sketch}. If this rotation angle was observed in
experiments, it would be a big challenge to the standard model of
particle physics. In next sections we will discuss how a
non-vanishing rotation angle appears in a Lifshitz-type
non-relativistic electrodynamics.

\section{Brief Introduction to a Lifshitz-type non-relativistic theory}

In this section we make a brief introduction to the Lifshitz-type
non-relativistic theory which is inspired by condensed matter
physics\cite{Horava:2008ih, Kluson:2009sm}. We work on a $d+1$
dimensional spacetime, with one time coordinate $t$ and $d$
spatial coordinates $x^i$. The theory of Lifshitz type exhibit
fixed points with anisotropic scaling governed by a dynamical
critical exponent $z$ as follows,
\begin{eqnarray}
t\rightarrow b^zt~,~~x^i\rightarrow b x^i~,
\end{eqnarray}
for an arbitrary constant $b$. In this case, time and space have
different dimensions with $[t]=-z$ and $[x^i]=-1$. Therefore, this
theory breaks a Lorentz symmetry at ultraviolet limit.

Suppose we have a covariant action $W$ in $d$ dimensional
Euclidean space for a general field $\psi^\alpha(\vec{x})$, then
it gives the partition function
\begin{eqnarray}
{\cal Z} = \int {\cal
D}\psi^{\alpha}(\vec{x})~e^{-W[\psi^{\alpha}(\vec{x})]}~.
\end{eqnarray}
In the standard quantum mechanism, this partition function
describes the norm of the ground-state functional, and so the
vacuum wave functional takes the form
\begin{eqnarray}
\Phi_g[\psi^{\alpha}(\vec{x})] = e^{-\frac{1}{2}W}~.
\end{eqnarray}
Further, we introduce the conjugate momentum in
Schr\"{o}dinger-type quantum field theory with canonical
commutation relation
\begin{eqnarray}
[\psi^{\alpha}(\vec{x}),\pi^{\beta}(\vec{y})]=i{\cal
G}^{\alpha\beta}\delta^3(\vec{x}-\vec{y})~,
\end{eqnarray}
where the matrix ${\cal G}$ is determined by the algebra in
internal space of the field. In the Schr\"{o}dinger
representation, this commutation relation requires the operator
form of the conjugate momentum to be equal to
\begin{eqnarray}
\pi_{\alpha}(\vec{x})=-i\frac{\delta}{\delta\psi^{\alpha}(\vec{x})}~.
\end{eqnarray}

Using the field $\psi^\alpha$ and its conjugate momentum
$\pi^\alpha$, one can write down the following operator,
\begin{eqnarray}\label{anni}
\hat{a}_{\alpha}(\vec{x}) \equiv i\pi_{\alpha}(\vec{x}) +
\frac{1}{2}\frac{\delta W}{\delta\psi^{\alpha}(\vec{x})}~.
\end{eqnarray}
If we use it to operate on the vacuum wave functional, it
annihilates the ground state
\begin{eqnarray}
\hat{a}_{\alpha} \Phi_g[\psi^{\alpha}(\vec{x})]=0~.
\end{eqnarray}
Thus it has a physical interpretation as an annihilating operator,
and correspondingly we can define its conjugate $\hat{a}^{\dag}$
as the creating operator.

The Hamiltonian of four dimensional spacetime can be constructed
by
\begin{eqnarray}
H = \int dx^3 {\cal H}(x) \equiv \int dx^3
\frac{1}{2}\hat{a}^{\dag}_\alpha{\cal
G}^{\alpha\beta}\hat{a}_\beta~.
\end{eqnarray}
Making use of the definition (\ref{anni}) and its conjugate, one
can obtain the Hamiltonian density
\begin{eqnarray}\label{hambasic}
{\cal H}(\vec{x}) = \frac{1}{2}\pi_\alpha{\cal
G}^{\alpha\beta}\pi_\beta +
\frac{1}{8}\frac{\delta{W}}{\delta\psi^\alpha} {\cal
G}^{\alpha\beta} \frac{\delta{W}}{\delta\psi^\beta}~,
\end{eqnarray}
Even in the classical Hamiltonian system, there is the equations
of motion
\begin{eqnarray}
D_t\psi^{\alpha}(\vec{x}) = \{\psi^{\alpha}(\vec{x}),H\} =
\pi^\alpha(\vec{x})~,
\end{eqnarray}
where $D_t$ is the time derivative which preserves
gauge-invariance.

Eventually, a Lagrangian density can be given by
\begin{eqnarray}\label{lagbasic}
{\cal L} &=& D_t\psi^{\alpha}\pi_\alpha - {\cal H}\nonumber\\
&=& \frac{1}{2}D_t\psi_{\alpha}{\cal
G}^{\alpha\beta}D_t\psi_{\beta} -
\frac{1}{8}\frac{\delta{W}}{\delta\psi^\alpha} {\cal
G}^{\alpha\beta} \frac{\delta{W}}{\delta\psi^\beta}~,
\end{eqnarray}
and consequently we can learn that in the Lifshitz-type theory a
Lagrangian density of a free field is a sum of kinetic term that
involves time derivative and a potential that is derived from the
three dimensional Euclidean action. Any model obtained through the
above process is said to exactly satisfy the {\it detailed balance
condition}.

One may keep in mind that, we can add some deformations to the
Lagrangian\cite{Horava:2008jf} if the relevant terms can keep the
gauge invariance, symmetry in internal space, and
renormalizability. These terms would bring a little violation of
the detailed balance condition, yet is still well-defined.

\section{Non-relativistic Electrodynamics Coupled with a scalar field in $d+1$ dimensional spacetime}

In this section we build a non-relativistic model describing a
scalar field $\phi$ coupled with a photon gauge field $A_\mu$ in
$d+1$ dimensional spacetime. Our model is required to be
 invariant under the following gauge transformation,
\begin{eqnarray}
 &\delta_\lambda A_t = {\dot\lambda}/{q}~,~~\delta_\lambda A_i =
{\partial_i\lambda}/{q}~,&
\end{eqnarray}
where $q$ is a coupling constant with its dimension to be $[q]=0$.
Therefore, the vector sector of the Lagrangian has to be
constructed by the field strengths $E_{i}$ and $F_{ij}$, which are
defined as follows,
\begin{eqnarray}
E_{i} = \dot A_i-\partial_iA_t~,~~ F_{ij} =
\partial_iA_j-\partial_jA_i~,
\end{eqnarray}
and the covariant derivative is given by
\begin{eqnarray}
D_t=\partial_t-iqA_t~,\nabla_i=\partial_i-iqA_i~.
\end{eqnarray}
Moreover, the engineering dimensions of the gauge field components
at the corresponding fixed point are:
\begin{eqnarray}
[A_t]=z~,~~[A_i]=1~.
\end{eqnarray}

Inspired by the structure of the Lifshitz-type theory, we take the
action with form of,
\begin{eqnarray}
S=\int dtdx^d [({\cal L}_K+{\cal L}_V)+{\cal L}_D]~,
\end{eqnarray}
which is constructed by kinetic part, potentials and some possible
deformations.

Using the notations of Ref. \cite{Horava:2008jf}, we give the
kinetic terms as follows,
\begin{eqnarray}\label{lk}
{\cal L}_K = \frac{1}{2}(\dot\phi)^2 + \frac{1}{2e^2}E_{i}E^{i}~,
\end{eqnarray}
with $e$ as a coupling constant with dimension
$[e]=1+\frac{z-d}{2}$.

The potential terms are constructed from the detailed balance
condition, which are given by,
\begin{eqnarray}
{\cal L}_V = -\frac{1}{8}(\frac{\delta
W}{\delta\phi})^2-\frac{e^2}{4}\frac{\delta{W}}{\delta
A_i}\frac{\delta{W}}{\delta A^i}~,
\end{eqnarray}
where $W[\phi,A^i]$ is the action in $d$ dimensional Euclidean
space as introduced in previous section. The advantage of the
detailed balance condition is that systems satisfy this condition
have simple quantum nature relevant to that of associated theory
described by the action $W$ in lower dimensions.

Without the violation of spatial rotation, we obtain the covariant
Euclidean action $W$ with an anisotropic scaling exponent up to
$z=2$ as follows,
\begin{eqnarray}
W[\phi,A_i] &=& \int dx^d \bigg[ (\sigma_1\phi\nabla^2\phi+
\sigma_n\phi^n)
\nonumber\\
&&+\frac{1}{2eg}(F_{ij}F^{ij}+m\epsilon^{ijk}A_i\partial_jA_k)
\bigg]~,
\end{eqnarray}
where $g$ is another coupling constant for gauge field with
dimension as $[g]=3-\frac{d+z}{2}$ and the last term in the
integral is the only relevant deformation preserving gauge
invariance, the coefficient $m$ has dimension unity. The
dimensions of the coupling constants for the scalar are
$[\sigma_1]=z-2$, and $[\sigma_n]=\frac{2-n}{2}d+\frac{n}{2}z$
respectively, where $n\geq2$ is an integer. According to Eq.
(\ref{lagbasic}), we take the functional derivative of the action
$W_A$ and finally obtain the potential terms
\begin{eqnarray}\label{lv}
{\cal L}_V^{\phi} &=&
-\frac{1}{2}\sigma_1^2(\nabla^2\phi)^2-\sigma_1\sigma_2\phi\nabla^2\phi-\frac{1}{2}\sigma_2^2\phi^2+...~,\\
{\cal L}_V^{A} &=&
-\frac{1}{2g^2}(\frac{m^2}{8}F_{ij}F^{ij}+\partial_jF^{ij}\partial^kF_{ik}
\nonumber\\ && +\frac{m}{2}\epsilon^{ijk}F_{jk}\partial^lF_{li})~.
\end{eqnarray}

To extend, we can add some deformations to the action if the
relevant terms can keep the gauge invariance and power-counting
renormalizability. These terms would bring a little violation of
the detailed balance condition, yet is still well-defined. For a
Maxwell field and a scalar field we considered in this letter,
there are three relevant terms of which the form take,
\begin{eqnarray}\label{Ldeform}
{\cal L}_D &=& -\frac{1}{g^2} (\kappa_0\epsilon^{ijk}A_iF_{jk}
+\kappa_1\dot\phi\epsilon^{ijk}A_iF_{jk} \nonumber\\
&& +\kappa_2\nabla_i\phi\epsilon^{ijk}A_jE_{k} )~,
\end{eqnarray}
which are Chern-Simons-like terms. The operators in Eq.
(\ref{Ldeform}) preserves the $U(1)$ gauge transformation and also
the shift symmetry $\phi\rightarrow\phi+C$ with $C$ a constant.
The requirement of the shift symmetry prohibits the large
radiative contribution to the potential of the scalar field
$\phi$\cite{Li:2001st}. The dimensions of the coefficients are
$[\kappa_0]=3$, and $[\kappa_1]=[\kappa_2]=3-\frac{d+z}{2}$
respectively. In the specific case we considered, there is $d=3$
and $z=2$ and thus we obtain $[\kappa_1]=[\kappa_2]=1/2$.
Therefore, these deformed terms have well-defined quantum
behaviors and so our model is power-counting renormalizable and
ghost-free. Note that, a power-counting renormalizability does not
necessarily imply the actual renormalizability. We will study this
issue in details in near future.

In low energy scale, this theory flows to $z=1$ and we expect the
Lorentz invariance can be recovered as an accidental symmetry. To
show that we rescale the time dimension,
\begin{eqnarray}
x^0=ct~,
\end{eqnarray}
then we redefine the scalar field,
\begin{eqnarray}
\tilde\phi=c^{\frac{1}{2}}\phi~,
\end{eqnarray}
and some relevant coefficients,
\begin{eqnarray}
\tilde\kappa_0=\frac{8}{m^3}\kappa_0~,
~\tilde\kappa_1=\frac{8c^{\frac{1}{2}}}{m}\kappa_1~,
~\tilde\kappa_2=\frac{8c^{\frac{1}{2}}}{m}\kappa_2~,
\end{eqnarray}
by the speed of light in terms of the ultraviolet variables
\begin{eqnarray}
c=\frac{em}{2g}~,
\end{eqnarray}
and use the relativistic notation $A_{\mu}=(A_t/c, A_i)$.

Consequently, if we focus on the gauge field sector and its
interaction with the scalar field, and work within $3+1$
dimensional spacetime, the Lagrangian becomes
\begin{eqnarray}\label{lagrangian}
\mathcal{L}^{A} &=& -\frac{1}{4\tilde{g}^2} \bigg[
F_{\mu\nu}F^{\mu\nu}
+ \frac{\tilde\kappa_1}{m}\tilde\phi{F}_{\mu\nu}\tilde{F}^{\mu\nu} \nonumber\\
&& +\frac{8}{m^2}\partial_jF^{ij}\partial^lF_{il}
+\frac{4}{m}\epsilon^{ijk}F_{jk}\partial^l F_{il} \nonumber\\
&& + 2\tilde\kappa_0m\epsilon^{ijk}A_iF_{jk} \bigg]~,
\end{eqnarray}
where we have required $\kappa_1=\kappa_2$ and used the signature
$(-,+,+,+)$ for the metric and the effective coupling constant is
\begin{eqnarray}
\tilde{g}^2=\frac{2eg}{m}~.
\end{eqnarray}
At the ultraviolet fixed point both $e$ and $g$ have dimensions of
$1/2$, so $\tilde{g}$ is dimensionless. We can see that the last
three terms in the Lagrangian (\ref{lagrangian}) violate the
Lorentz invariance explicitly while the send one violates the
Lorentz invariance dynamically if $\tilde{\phi}$ develops a
non-zero time derivative which has been obtained by supposing a
scalar field couples to an anomalous $B-L$ current as observed by
Ref. \cite{Li:2006ss}. Moreover, the possible couplings of
$\tilde\phi$ to the standard model particles induce observable
effects and have interesting implications in particle
physics\cite{Li:2001st, DeFelice:2002ir, Kostelecky:2002ca,
Geng:2007ga} and cosmology\cite{Davoudiasl:2004gf, Li:2004hh,
Mukhopadhyay:2007vca} (see also \cite{Bertolami:1996cq}). However,
since no definite signal of these CPT violating effects has been
observed in the ground-based laboratories, we would like to see if
they could leave signals on cosmological observations
\cite{Carroll:1989vb, Carroll:1991zs, Carroll:1997tc}.

\section{The propagation of the photon and the rotation angle}

In this section we will study the propagation of the photon
described by the Lagrangian in the infrared limit
(\ref{lagrangian}). Because $A_0$ is non-dynamical and this theory
is gauge invariant, there are only two dynamical degrees of
freedom for the photon. It is natural to choose the Coulomb gauge
$\partial_i A^i=0$ to solve the equation of motion due to the
non-relativistic nature. In the absence of other source, we can
set $A_0=0$. Besides, we require the scalar field is homogeneous
and is only the function of time $\tilde\phi=\tilde\phi(t)$(we
refer to \cite{Li:2008tma, Pospelov:2008gg, Kamionkowski:2008fp,
Yadav:2009eb, Wang:2009mj, Gluscevic:2009mm} for an inhomogeneous
rotation).

So, the equation of motion followed from (\ref{lagrangian}) is
\begin{eqnarray}
\partial_0^2 A_i -\nabla^2A_i +\frac{4}{m^2}\nabla^4A_i +2(\tilde\kappa_0m
+\frac{\tilde\kappa_1}{m}\partial_0\tilde\phi)\epsilon^{ijk}\partial_jA_k
\nonumber\\ -\frac{4}{m}\epsilon^{ijk}\partial_j\nabla^2 A_k=0~.
\end{eqnarray}
With the ansatz of plane wave $A_i=a_ie^{ik_{\rho}x^{\rho}}$, and
assume the wave is propagating along the $+z$ direction of
coordinate with the wave vector $k_{\rho}=(\omega,~0,~0,~k)$, we
soon get $a_3=0$ from the gauge condition and the equations for
the dynamical components:
\begin{eqnarray}
&(-\omega^2+k^2+\frac{4k^4}{m^2})a_1 - i[2k(\tilde\kappa_0m
+\frac{\tilde\kappa_1}{m}\partial_0\tilde\phi)+\frac{4k^3}{m}]a_2=0& \nonumber\\
&(-\omega^2+k^2+\frac{4k^4}{m^2})a_2 + i[2k(\tilde\kappa_0m
+\frac{\tilde\kappa_1}{m}\partial_0\tilde\phi)+\frac{4k^3}{m}]a_1=0&. \nonumber\\
\end{eqnarray}
These are algebraic equations. The nontrivial solution requires
the matrix of the coefficients has null determinant. This
requirement is explicitly
\begin{eqnarray}
(-\omega^2+k^2+\frac{4k^4}{m^2})^2 = [2k(\tilde\kappa_0m
+\frac{\tilde\kappa_1}{m}\partial_0\tilde\phi)+\frac{4k^3}{m}]^2~,
\end{eqnarray}
and so yields,
\begin{eqnarray}\label{dispersion}
\omega^2_{\pm} = k^2(1 \mp \frac{2}{m} k)^2 \mp 2k(\tilde\kappa_0m
+\frac{\tilde\kappa_1}{m}\partial_0\tilde\phi)~.
\end{eqnarray}
This modified dispersion relation leads to energy-dependent group
velocities and will induce the time-delay of
photons\cite{Ellis:1999sd, Boggs:2003kxa}, which might be an
observable effect in Gamma ray burst
experiments\cite{Abdo:2009zz}.

We can see from above equation that the dispersion relations for
left- and right-handed circular polarized components of photons
are different. The polarization angle of the linear polarized
photon will be rotated by an angle $\Delta\chi$ after a period of
propagation\cite{Harari:1992ea, Lue:1998mq}.

In the Friedmann-Robertson-Walker (FRW) universe, the rotation
angle at high frequency regime ($k_{ph}$ is quite large) is given
by
\begin{eqnarray}
\Delta\chi_{UV} &=& \int_{z_i}^0 (\omega_--\omega_+)dt(z') \nonumber\\
&\simeq& -\frac{4}{m}\int_{z_i}^0
\frac{k_{ph}^2(z')dz'}{H(z')(1+z')}~,
\end{eqnarray}
where $k_{ph}(z')=(1+z')k_c$ is a physical frequency and $z_i$ is
the redshift when the photons were emitted. For the CMB, the
redshift is about $1100$. However, if $k_{ph}$ has already lied in
the low frequency regime when the photons start propagations, the
rotation angle can be derived by making use of geometric optical
approximation as generally discussed in \cite{Li:2008tma},
\begin{eqnarray}\label{chiIR}
\Delta\chi_{IR} &\simeq& \int_{z_i}^0 2(\tilde\kappa_0m
+\frac{\tilde\kappa_1}{m}\partial_0\tilde\phi)dt(z') \nonumber\\
&\simeq& \int_{z_i}^0 \frac{2(\tilde\kappa_0m
+\frac{\tilde\kappa_1}{m}\partial_0\tilde\phi)}{H(z')(1+z')}dz'~.
\end{eqnarray}
This result is similar to the case of cosmological CPT violation
as considered in Ref. \cite{Carroll:1989vb}, and the current CMB
data mildly indicates a non-zero central value\cite{Feng:2006dp,
Komatsu:2008hk, Xia:2008si, Kostelecky:2008be}.

Notice that, this rotation angle is frequency dependent for high
frequency photons, different from that of low energy photons. This
frequency dependence is different from that of Faraday rotation in
which the polarization of the photon is rotated when it pass
through a magnetic field. In the later case, the rotation angle is
inversely proportional to the square of the frequency of the
photon, in the case here however it is proportional to the square
of the frequency. We note that similar frequency dependent
rotation angle was studied in Ref. \cite{Kostelecky:2007zz} and
recently in Ref. \cite{Gubitosi:2009eu} from the phenomenology of
quantum gravity \cite{Myers:2003fd}.

\begin{figure}[htbp]
\includegraphics[scale=0.9]{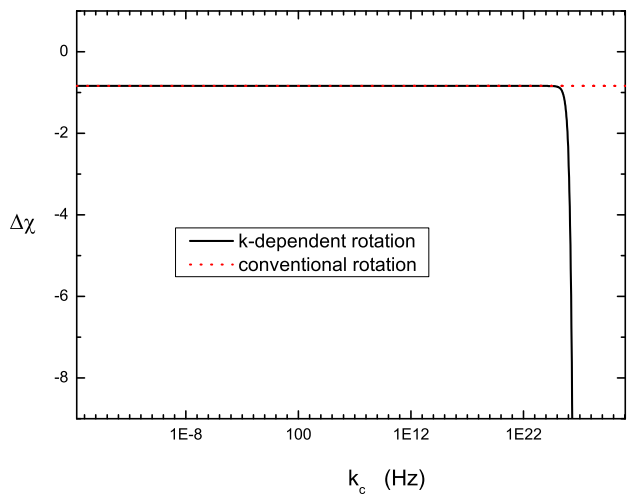}
\caption{Plot of the rotation angle $\Delta\chi$ as a function of
the comoving wavenumber $k_c$. The black solid line represents the
rotation angle in the non-relativistic Maxwell theory, and the red
dot line represents the conventional one. In the numerical
calculation, we take the following parameters
$\tilde\kappa_0=10^{-52},~m=10^{49}{\rm Mpc}^{-1}$, and use the
WMAP5 results\cite{Komatsu:2008hk}
$H_0=70.5kms^{-1}Mpc^{-1},~\Omega_{m0}=0.2736,~\Omega_{\Lambda0}=0.726$,
and $w_{\Lambda}=-1$ is assumed.} \label{fig:angle}
\end{figure}

An important issue concerns that, if our model can be detected or
constrained by current and future observations. From the above
results, we can see that it is directly determined by the value of
the non-relativistic scale $m$ appeared in the Lagrangian. To be
precise, a general extension of Maxwell theory with various
Lorentz violating terms has been studied by Kostelecky and Mewes
in Refs. \cite{Kostelecky:2007zz, Kostelecky:2008be}. The third
and the forth terms of Eq. (\ref{lagrangian}) in our model
correspond to the $k^{(5)}$ and $k^{(4)}$ ones appeared in
\cite{Kostelecky:2007zz, Kostelecky:2008be} respectively, if we
fix $\partial_\mu\tilde\phi$ to be constant. Therefore, by virtue
of the results obtained in Refs. \cite{Kostelecky:2007zz,
Kostelecky:2008be}, we conclude that the non-relativistic scale
$m$ approximately lies at the level of $10^{11}{\rm
GeV}$\footnote{There is roughly $m\sim k^{(4)}/k^{(5)}$ where
$k^{(4)}\sim 10^{-31}$ and $k^{(4)}\sim 10^{-20}{\rm GeV}^{-1}$
from \cite{Kostelecky:2007zz}.}. Then we study how this parameter
affect the rotation angle at high energy scale. As a first glance,
we neglect the contribution of the rolling scalar $\phi$ and focus
on the IR limit. The current CMB polarization constraints require
the term $\tilde\kappa_0m$ is of order $O(10^{-3}{\rm Mpc}^{-1})$.
In order to let B-mode signal from the UV limit be observable, the
value of mass scale $m$ has to be promoted. Specifically, we
consider the case with $m=10^{49}{\rm Mpc}^{-1}(\sim10^{11}{\rm
GeV})$ which also satisfies the constraints from Ref.
\cite{Kostelecky:2007zz} and do the numerical calculations without
any approximations, as shown in Figure \ref{fig:angle}. The result
shows that the rotation angle coincides with the conventional one
in Eq. (\ref{chiIR}) at low energy scale, but becomes
frequency-dependent and negative at high energy limit. From the
recent CMB polarization observations, namely QUaD
data\cite{Hinderks:2008yg}, a negative frequency-dependent
rotation angle might be favored at high energy
scale\cite{Brown:2009uy, Xia:2009ah}. It deserve a fully detailed
analysis on constraining our model with current and future
cosmological observations. We will work on this issue in near
future studies.

\section{Conclusion and Discussions}

The philosophy of a Lifthitz-type non-relativistic theory with
quantum criticality was motivated by a quantum theory of multiple
membranes designed such that the ground-state wavefunction of the
membrane with compact spatial topology reproduces the partition
function of the bosonic string on worldsheet\footnote{Note that
the scenario of multiple branes has been recently used to drive
inflation as studied in Refs. \cite{Cai:2008if, Cai:2009hw}.}. Due
to its advantage on power-counting renormalization and
non-relativistic peculiarity, this theory has been awoken and
provided plentiful phenomenons when applied in various fields and
so attracted a lot of attentions. For example as pointed out by
Refs. \cite{Brandenberger:2009yt, Cai:2009in}, because of its
non-relativistic feature, the gravity sector in the Lifshitz-type
theory could effectively violate energy conditions and so can lead
to a solution of bouncing universe \cite{Cai:2007qw} and the
cosmological perturbations in low energy regime form a
scale-invariant spectrum \cite{Cai:2007zv, Cai:2008qb, Cai:2008qw}
with a potential red tilt\cite{Cai:2009hc}. However, before
phenomenological applying this theory into any subjects, we ought
to consider whether this theory is discernible for observations.

In this letter we have studied a non-relativistic model for a
gauge field coupled with a scalar field, based on the construction
of the Lifshitz-type theory. Specifically, we considered a model
of electrodynamics which exhibits anisotropic scaling between time
and space with the dynamical critical exponent $z=2$. The theory
is non-relativistic in essence, and hence the Lorentz invariance
is violated and can only be recovered in the low energy regime as
an approximate symmetry by accident. Accompanied with the
violation of Lorentz invariance, the CPT symmetry is also broken
due to an existence of some Chern-Simons-like terms. This CPT
violation can be viewed as the Trans-Planckian
physics\cite{Brandenberger:1999sw, Martin:2000xs} and lead to the
generations of $TB$ and $EB$ modes on CMB polarization spectra and
a rotation of the polarization of the photon propagated in the
universe\footnote{The similar CPT breaking terms can also appear
in the gravity sector of Ho\v{r}ava-Lifshitz
theory\cite{Takahashi:2009wc} which may affect CMB tensor
spectrum\cite{Cai:2007xr}.}, and the rotation angle is frequency
dependent in UV limit while similar to the case of a dynamical
violation in IR limit. This signature, if detected by future CMB
observations with higher precision, would be an important clue to
new physics beyond standard model.

\begin{acknowledgments}
We would like to thank Robert Brandenberger, Bin Chen, Jie Liu,
and Yun-Song Piao for discussions and comments on the manuscript.
The author M.L. is supported in part by National Science
Foundation of China under Grants No. 10575050 and No. 10775069.
The research of Y.C. and X.Z. is supported in part by the National
Science Foundation of China under Grants No. 10533010 and
10675136, and by the Chinese Academy of Sciences under Grant No.
KJCX3-SYW-N2.
\end{acknowledgments}

\end{document}